\documentclass{elsart}
\usepackage{graphicx}

\begin{document}

\graphicspath{
  {/u/shikaze/work/paper/eps/}
  }
\DeclareGraphicsRule{.ps.gz}{eps}{.ps.bb}{`/usr/local/bin/zcat #1}
\DeclareGraphicsRule{.eps.gz}{eps}{.eps.bb}{`/usr/local/bin/zcat #1}

\begin{frontmatter}
  \title{
    Large-Area Scintillator Hodoscope with\\
    50~ps Timing Resolution Onboard BESS
  }
  \author[Tokyo]{Y.~Shikaze}\footnotemark[1],
  \author[Tokyo]{S.~Orito},
  \author[Kobe]{T.~Mitsui},
  \author[Tokyo]{K.~Yoshimura},
  \author[Kobe]{H.~Matsumoto},
  \author[Tokyo]{H.~Matsunaga},
  \author[Kobe]{M.~Nozaki},
  \author[Tokyo]{T.~Sonoda},
  \author[Tokyo]{I.~Ueda},
  \author[KEK]{T.~Yoshida}
  \address[Tokyo]{
    Department of Physics and International Center for Elementary Particle 
    Physics, University of Tokyo, Hongo, Bunkyo, Tokyo, 113-0033 Japan
    }
  \address[Kobe]{
    Department of Physics and Graduate School of Science and Technology, 
    Kobe University, Rokkodai-cho, Nada, Kobe, 657-8501 Japan
    }
  \address[KEK]{
    High Energy Accelerator Research Organization (KEK), 
    Oho, Tsukuba, Ibaraki, 305-0801 Japan
    }
  \footnotetext[1]{Corresponding author. Tel.: +81 3 3815 8384;fax: +81 3 3814 8806;e-mail: shikaze@icepp.s.u-tokyo.ac.jp.}
  \begin{abstract}
 We describe the design and performance of a large-area scintillator hodoscope 
onboard the BESS rigidity spectrometer; an instrument with an acceptance 
of 0.3~m$^{2}$sr.
 The hodoscope is configured such that 10 and 12 counters are respectively 
situated in upper and lower layers.
 Each counter is viewed from its ends 
by 2.5 inch fine-mesh photomultiplier tubes placed in a stray magnetic 
field of 0.2~Tesla.
 Various beam-test data are presented.
 Use of cosmic-ray muons at ground-level confirmed 50~ps timing resolution 
for each layer, giving an overall time-of-flight resolution of 70~ps rms 
using a pure Gaussian resolution function.
 Comparison with previous measurements on a similar scintillator hodoscope
indicates good agreement with the scaling law that 
timing resolution is proportional to 1/$\sqrt{N_{\rm pe}}$, 
where $N_{\rm pe}$ is the effective number of photoelectrons.
  \end{abstract}
\end{frontmatter}

\section{Introduction}
BESS is a balloon-borne rigidity spectrometer having an acceptance 
of 0.3~m$^{2}$sr designed~\cite{kn:prop87} as an omni-purpose 
spectrometer with special emphasis given to the measurement of cosmic-ray 
antiprotons.
 In three consecutive scientific flights over northern Canada (1993--1995), 
we measured the rigidity ($R \equiv p/Z$, momentum per charge), velocity, 
and d$E$/d$x$ of about 12 million cosmic-ray particles transversing through it.
 The first mass-identified, and thus unambiguous, detection of cosmic-ray 
antiprotons was achieved by BESS'93 in the kinetic energy range from 
0.3 to 0.6~GeV~\cite{kn:pbar93}.
 In 1995, time-of-flight (TOF) resolution was improved from 280 to 
110~ps, which led to background-free detection of 43 antiprotons with 
energies from 0.18 to 1.4~GeV~\cite{kn:pbar95}.

When we search for low energy ``primary'' antiprotons from novel 
sources~\cite{kn:pbh}, 
the statistical accuracy must be improved and the 
energy range of antiproton identification enlarged so that the expected 
peak of ``secondary'' antiprotons at 2~GeV, i.e., those produced by collisions
of high energy cosmic-rays with Galactic interstellar medium, can be detected 
and its absolute flux measured.
 Independent of the search for the novel primary antiproton component, precise
measurement of the secondary antiprotons is in itself crucially important 
both to understand solar modulation and determine the propagation mechanism of
cosmic-rays in the Galaxy.

 Accordingly, we proposed~\cite{kn:prj} obtaining background-free, positive 
(mass-identified) detection of antiprotons up to 3~GeV by improving the timing
resolution of TOF counters to 50~ps~\cite{kn:tof} and eliminating the 
overwhelming $e^- /\mu^- $ background with a threshold-type aerogel 
\v{C}erenkov counter having a refractive index of 1.02 to 1.03.
 A reduction in timing resolution is also necessary to identify light
isotopes in cosmic-rays.

 Having described the performance of the aerogel \v{C}erenkov 
counter~\cite{kn:acc}, here we present the design, construction, 
and performance of the enhanced TOF system.

\section{Spectrometer}
\begin{figure}[hbtp]
  \begin{center}
    \includegraphics[width=9cm]{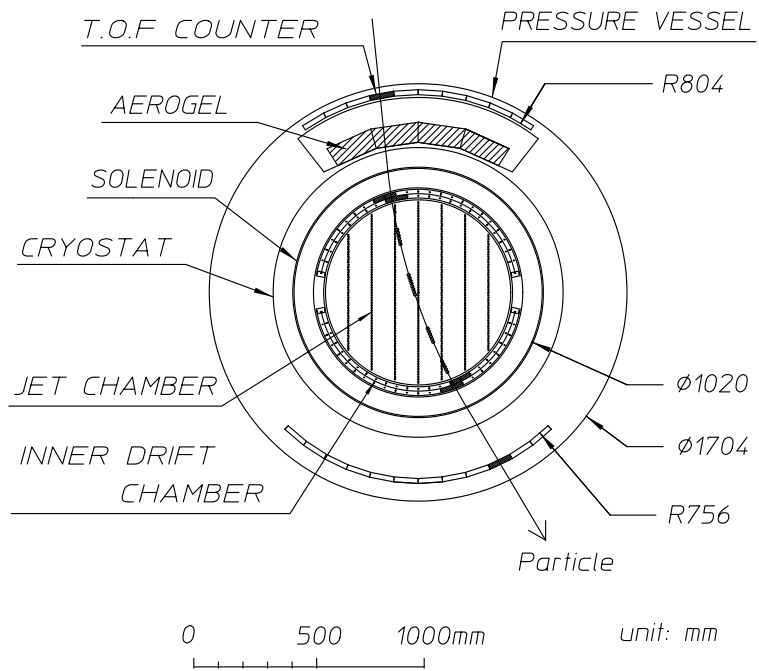}
  \end{center}
  \caption
  [Cross-section of the BESS'97 spectrometer.]
  {Cross-section of the BESS'97 spectrometer 
   showing a positive-charged particle event.}
\label{fig:bess}
\end{figure}

Figure~\ref{fig:bess} shows a cross-section of the BESS rigidity 
spectrometer~\cite{kn:pbar93,kn:det} in its '97 configuration.
 The spectrometer consists of a superconducting solenoidal magnet, inner 
drift chambers (IDCs), a JET drift chamber, a TOF plastic scintillation 
hodoscope, and an aerogel \v{C}erenkov counter.
 A uniform field of 1~Tesla is produced by a thin (4 g/cm$^2$) superconducting 
coil~\cite{kn:mag} which allows particles to pass through with small 
interactions.
 The magnetic field region is filled with the central tracking volume.
 The employed geometry gives an acceptance of 0.3~m$^2$sr, being an order of 
magnitude larger than that of previous cosmic-ray spectrometers.
 Tracking is performed by fitting up to 28 hit-points in the drift chambers, 
resulting in a magnetic-rigidity resolution of 0.5~\% at 1~GV/$c$.
 This continuous, redundant, three-dimensional tracking enables recognizing
multi-track events and tracks having interactions or scattering: a
feature minimizing such background.
 The upper and lower scintillators provide two independent d$E$/d$x$ 
measurements and the TOF of particles.
 We also measure d$E$/d$x$ in the drift chamber gas obtained as a 
truncated mean of the integrated charges of hit pulses.

 Unlike the BESS'95 configuration in which two layers of TOF counters were 
both cylindrically arranged at a radius of 660~mm, in the new arrangement 
the upper and lower layers have a respective radius of 804 and 756~mm.
 The extension of path length from upper to lower layer and the 
upgraded timing resolution improve velocity measurements.

\section{TOF hodoscope and PMT}
\begin{figure}[hbtp]
  \begin{center}
    \includegraphics[width=10cm]{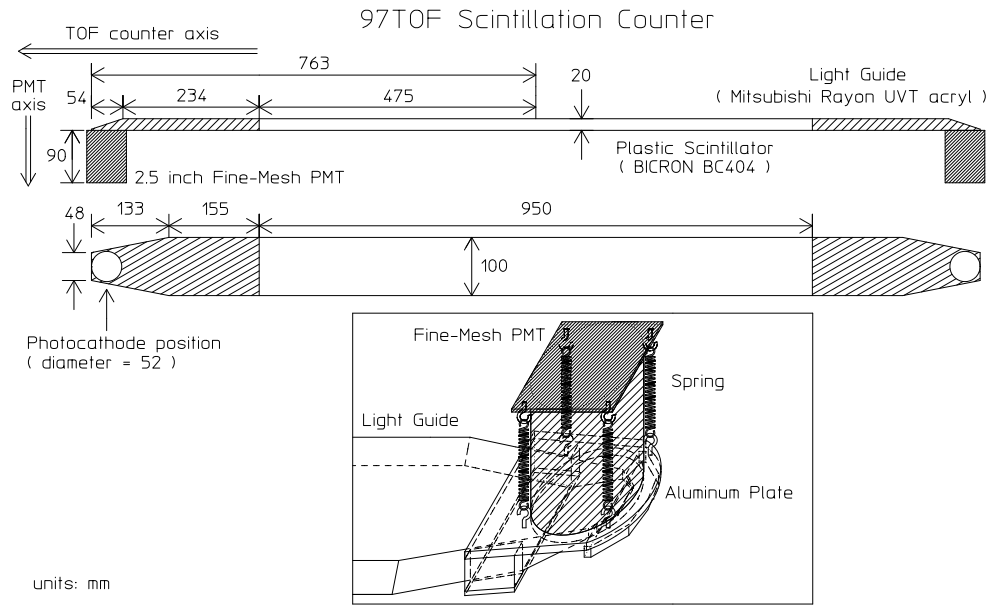}
  \end{center}
  \caption
  [Overview of a TOF counter for BESS'97.]
  {Overview of BESS'97 TOF counter incorporating 2.5 inch FM-PMTs.}
  \label{fig:lg3c}
\end{figure}

\begin{figure}[hbtp]
  \begin{center}
    \includegraphics[width=6cm]{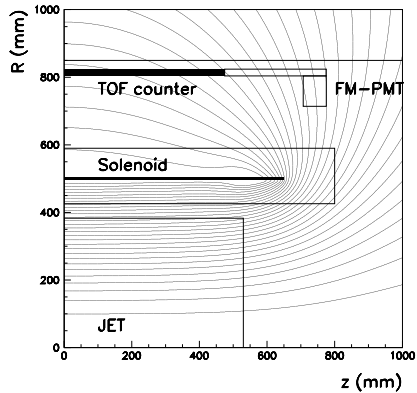}
  \end{center}
  \caption
  [Flux line of the magnetic field around the FM-PMT.]
  {Flux line of the magnetic field around the FM-PMT.}
  \label{fig:bfield4}
\end{figure}

 The TOF hodoscope consists of 10 upper and 12 lower plastic scintillation 
counter paddles (950 $\times$ 100 $\times$ 20~mm$^{3}$, BICRON BC-404, 
refractive index $n$ = 1.58).
 A light guide (Fig.~\ref{fig:lg3c}), made of a UV-transparent acryl 
plate (Mitsubishi Rayon, $n$ = 1.52), 
is glued with optical epoxy (BICRON BC-600, $n$ = 1.56) to 
each end of the scintillator and connected to a 2.5 inch fine-mesh (FM) 
magnetic-field-resistant photomultiplier tube (PMT) (Hamamatsu R6504S) 
having an effective photocathode diameter of 52~mm.
 To minimize loss of photoelectrons due to magnetic effects, PMTs are placed 
tangential to the light guide 
with the result that the angle between PMT axis and 
magnetic field lines is less than $16^{\circ}$ (Fig.~\ref{fig:bfield4}).
 For a nominal central magnetic field of 1~Tesla, the stray field at 
PMT positions is 0.2~Tesla.
PMT ends with attached light 
guide are fit into circular holes in an aluminum plate.
 Four springs (Fig.~\ref{fig:lg3c}) are used to prevent separation from the 
light guide.
 To minimize shock and vibration during shipping, launching, and landing, 
a 1-mm-thick silicon pad (Shin-Etsu OF113, $n$ = 1.51) 
is sandwiched between the light 
guide and PMT.
 Both sides of the pad are coated with optical grease (Shin-Etsu Optseal, 
$n$ = 1.47).


 The most suitable shape of the light guide with regard to timing resolution 
and effective~\cite{kn:eff} number of 
photoelectrons ($N_{\rm pe}$) was determined by investigating various 
fish-tail-shaped light guides using 
a simulation code (GUIDE7 ~\cite{kn:gd7}) and cosmic-ray and 
beam test data.
 After selecting three candidates in which the shape of the planes were 
optimized based on simulations and cosmic-ray data, the best was determined 
using the beam test data.

 In the '95 flight, we individually wrapped scintillators with a white 
sheet (Millipore IPVH00010, white filter paper) for diffusive reflection of 
light.
 As the sheet ended up partially sticking to the scintillator, the light 
collection efficiency might have been reduced.
 Accordingly, other wrapping materials were considered and aluminized Mylar 
was selected.
 Scintillators and light guide are respectively wrapped in a 50- and 
25-$\mu$m-thick sheet of aluminized Mylar, and then collectively wrapped 
in a 200-$\mu$m-thick sheet of black vinyl for light shield.

\begin{figure}[hbtp]
  \begin{center}
    \includegraphics[width=8cm, height=12cm, angle=-90]{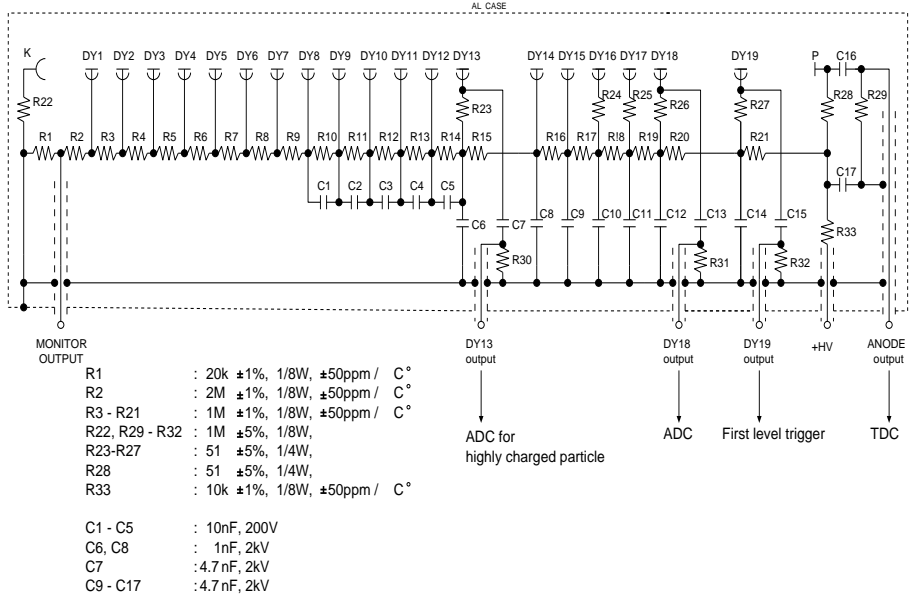}
  \end{center}
  \caption
  [Circuit diagram of the FM-PMT Hamamatsu R6504S voltage divider.]
  {Circuit diagram of the FM-PMT Hamamatsu R6504S voltage divider.}
  \label{fig:pmdiv}
\end{figure}

 The PMTs have a bialkali (Sb-Rb-Cs, Sb-K-Cs) photocathode whose effective 
diameter is typically 52~mm, being about 2.0 times larger area 
compared to that of the 2 inch FM-PMT used in the '95 flight.
 They exhibit a wide sensitivity for wavelengths from 300 to 650~nm, 
which well matches the scintillator light spectrum having maximum emission 
at 408~nm.
 The PMT has 19 dynodes which are situated parallel to the next dynode 
and separated about 0.8~mm apart.
 Electrons are accelerated by parallel electric fields between the dynodes; 
hence allowing the device to be used in a strong magnetic field as far as the 
direction of the magnetic field (0.2~Tesla) 
is parallel to the PMTs longitudinal axis.
 Figure~\ref{fig:pmdiv} shows a circuit diagram of the voltage divider in 
which the signal from the anode provides timing information and that of the 
19th dynode is used as a first-level trigger.
 Signals from the 13th and 18th dynodes are used 
for the charge measurement.
 Though the signal of FM-PMT is generally difficult to saturate, 
highly charged particles in the cosmic-ray experiment generate 
the signal with huge charge and cause the saturation of the signal.
 The signal from the 13th dynode is used for highly charged particles since 
it is difficult to saturate compared to the 18th one.

 As indicated in Fig.~\ref{fig:pmdiv}, 
capacitors (C8--C12,C14) connect dynodes following the 13th and ground.
 These capacitors serve to reduce crosstalk of the 13th dynode, which is 
affected by the large charge in cascading dynodes due to highly 
charged particles.
 Accordingly, the signal from 13th dynode is available for these particles.
 Its performance was tested using a blue LED (NICHIA,NLPB) in which light 
input to PMT was reduced by an optical filter (Fuji-film, ND filter) 
while the light output of the LED was constant.

\begin{figure}[hbtp]
  \begin{center}
    \includegraphics[width=8cm]{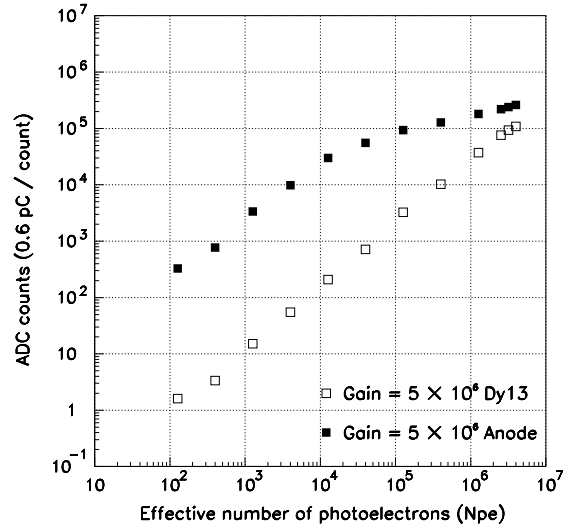}
  \end{center}
  \caption
  [Mean ADC counts versus $N_{\rm pe}$ for the anode and 13th dynode, 
   respectively.]
  {Mean ADC counts versus $N_{\rm pe}$ for the anode and 13th dynode, 
   respectively.}
  \label{fig:satu}
\end{figure}

Figure~\ref{fig:satu} shows analog-to-digital converter (ADC) counts 
versus $N_{\rm pe}$, i.e., light 
input illumination to the PMT, for both the anode and the 13th dynode, 
where the 13th dynode demonstrates good linearity and no saturation up to the 
$N_{\rm pe}$ = 4 $\times$ $10^{6}$; a value 
corresponding to $10^{4}$ times the light output of minimum ionizing 
particles.
 $N_{\rm pe}$ is evaluated as 
$\left( \langle ADC \rangle / \sigma \right)^{2}$, 
where $\langle$ ADC $\rangle$ is mean ADC counts 
and $\sigma$ the rms of the ADC distribution.

\section{Principle of timing measurement}
 We discuss here on the crossing time of a particle and 
time-of-flight, i.e., their difference between upper and lower layers of 
TOF counters of the BESS spectrometer.

 PMT signals have time jitter associated with pulse height variation, 
the so called time-walk effect~\cite{kn:tw}.
 The crossing time measured by time-to-digital converter (TDC) must 
therefore be corrected for this effect.
 The time-walk corrected timing for PMT $i$, $t_{ic}$, can be well fitted with 
\begin{equation}
  t_{ic} = t_{i} - W_{i}/\sqrt{q_{i}} ,
  \label{eq:slew1}
\end{equation}
where $t_{i}$, $q_{i}$, and $W_{i}$ are respectively the measured TDC time, 
measured integrated charge of the PMT signal, and a correction parameter 
determined by data fitting.
Subscripts 1 and 2 are used to denote PMTs at each end of a TOF counter.

\begin{figure}[hbtp]
  \begin{center}
    \includegraphics[width=6cm, angle=-90]{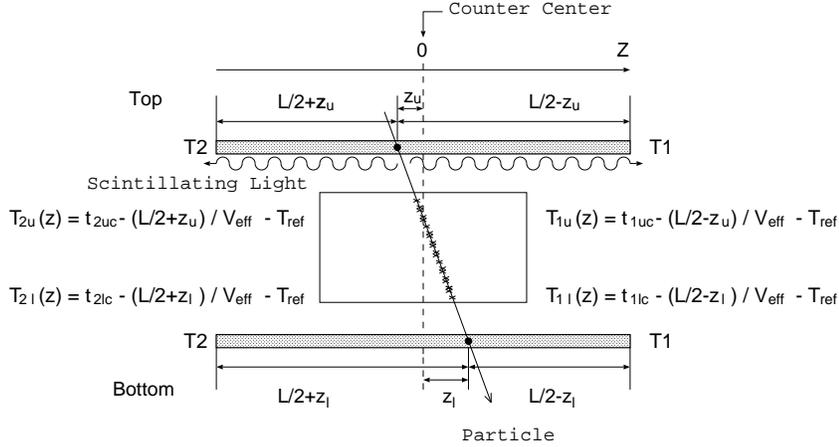}
  \end{center}
  \caption
  [Diagrammatic representation showing how crossing time is determined.]
  {Diagrammatic representation showing how crossing time is determined.}
  \label{fig:zzz}
\end{figure}

 Using the time-walk corrected timing for each PMT, we can calculate the 
crossing time based on the hit position and timing information.
 The hit position of a counter is defined using the $z$-axis along the 
counter's longitudinal direction (Fig.~\ref{fig:zzz}) in which the center of 
the counter is considered $z = 0$.
 Extrapolation of the trajectory determined by the central 
trackers (JET, IDCs) with good precision enables tracing back exactly to 
the impact points ($z$ and $r$-$\phi$ position) on the TOF counters.
 Crossing time of a particle is measured as the time difference between the 
reference time, $T_{\it ref}$, taken as the TDC common start time, and the 
time when a particle crosses the TOF counter.
 That is, for PMT $i$, $T_{i}(z)$ are
\begin{equation}
  T_{i}(z) = t_{ic} - (\frac{L}{2}+(-1)^{i}z)/V_{\it eff} - T_{\it ref} ,
  \label{eq:zcl}
\end{equation}
where $t_{ic}$ and $T_{\it ref}$ are the time-walk corrected timing and 
reference time, while $z$ is the hit position of the counter, $L$ the length 
of the counter including light guides, and $V_{\it eff}$ the effective 
velocity of light in the scintillator.

 Crossing time of a TOF counter is then obtained using $T_1(z)$ and $T_2(z)$ 
in which the weighted average of these crossing time 
measurements, $T_{w.a.}(z)$, is
  \begin{equation}
    T_{w.a}(z) \equiv \frac{T_{1}(z)/{{\sigma_{1}}^{2}(z)} + T_{2}(z)/{{\sigma_{2}}^{2}(z)}}{1/{{\sigma_{1}}^{2}(z)} + 1/{{\sigma_{2}}^{2}(z)}} ,
  \label{eq:pd3}
  \end{equation}
where $\sigma_{1}(z)$ and $\sigma_{2}(z)$ are the timing resolution of 
$t_{1c}$, $t_{2c}$ as a function of $z$, respectively.
 Using the crossing time for upper and lower TOF counters, $T_{U}$ and 
$T_{L}$, which are determined by Eq. (\ref{eq:pd3}), the TOF of the BESS 
spectrometer, $T_{\it tof}$, is expressed as $T_{\it tof} = T_{L} - T_{U}$.
 It is important to note that in the equation on $T_{\it tof}$, 
since $T_{\it ref}$ in $T_{U}$ and $T_{L}$ cancel out, 
time jitter regarding reference time has no effect on the resolution of 
$T_{\it tof}$.
\section{Beam test performance}
\subsection{Beam line}
\begin{figure}[hbtp]
  \begin{center}
    \includegraphics[width=10cm]{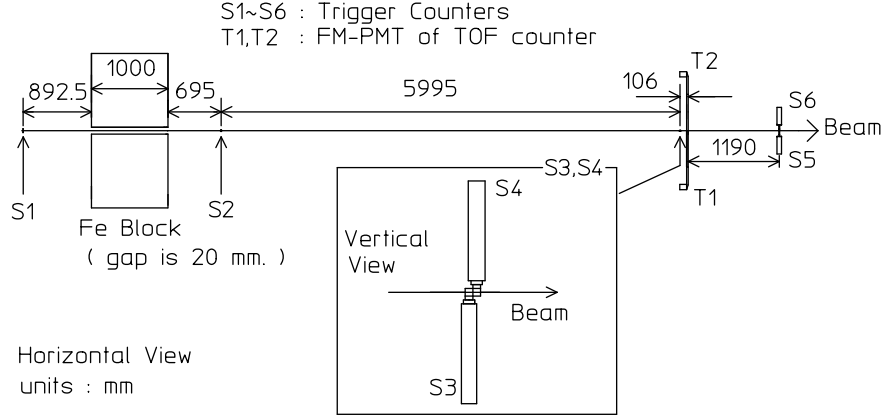}
  \end{center}
  \caption
  [Top and side views of beam testing setup.]
  {Top and side views of beam testing setup.}
  \label{fig:bm97feb}
\end{figure}

 Performance of the TOF counter was evaluated by beam test under no 
magnetic field.
 Beam tests were performed at KEK in $\pi$-2 beam line using 1~GeV/$c$ 
proton($p$)/$\pi^{+}$ beams and a 4~GeV/$c$ $\pi^{-}$ beam.
 Figure~\ref{fig:bm97feb} shows the employed setup, where S1--S6 are trigger 
counters that provide data used for $\pi^{+}$/$p$ separation 
by TOF measurement.
 S3 and S4 are reference counters 
(5(w) $\times$ 10(h) $\times$ 10(t)~mm$^{3}$) 
which generate TDC start and determine beam position.
 T1 and T2 are FM-PMTs of the TOF counter, each having an anode pulse height 
adjusted to 2~V for minimum ionizating $\pi$'s vertically incident to the 
counter center.
 For the data-taking, the CAMAC system was employed through the beam test.
ADC (LeCroy 2249W, 0.25~pC/count), TDC (Phillips 7186, 25~ps/count), and 
discriminator (Phillips 708, 300~MHz) were used as the CAMAC modules.
 The optimal threshold to timing measurement was found to be 20~mV to obtain 
best timing resolution.
 Position dependence of counter performance was investigated by obtaining 19 
data points along the $z$-direction (Fig.~\ref{fig:zzz}) at 50~mm intervals.

\subsection{Performance}
\subsubsection{Timing resolution}
 In the analysis of beam test data, the timings of the reference
counters, S3 and S4, are used as the reference time $T_{\it ref}$.
 The position of the reference
counters defines the beam position with an accuracy of 
$\pm2.5$~mm.
 These values are used to determine $T_{1}(z)$ and $T_{2}(z)$ 
in Eqs. (\ref{eq:zcl}).
 Because the reference counters are located very close to the test counter 
(Fig.~\ref{fig:bm97feb}), the jitter caused by the change of the beam status 
during passing through these counters is negligibly small 
for the beam crossing times of these counters.
 Accordingly, there is a relationship among the measured rms of $T_{i}$ and 
the timing resolution of $t_{ic}$ and the timing resolution of 
$T_{\it ref}$, i.e.,
\begin{equation}
    {\sigma_{T_{i}}}^{2} = {\sigma_{i}}^{2} + {\sigma_{\it ref}}^{2} 
                                            + {\sigma_{pos}}^{2} ,
\end{equation}
where $\sigma_{T_{i}}$, $\sigma_{i}$, $\sigma_{\it ref}$, and $\sigma_{pos}$ 
are respectively 
the measured rms of $T_{i}$, timing resolution of $t_{ic}$, timing 
resolution of $T_{\it ref}$, and the time jitter due to beam width.
 The resolution of the reference time is estimated as described next.
 Considering the reference time to be the mean time of adjacent trigger 
counters S3 and S4 (Fig.~\ref{fig:bm97feb}), the time-walk corrected timing 
for S3 and S4 ($T_{S3}$ and $T_{S4}$) gives $T_{\it ref}=(T_{S3}+T_{S4})/2$.
 The rms of the distribution of $(T_{S3}-T_{S4})/2$ is then used to estimate 
the timing resolution of $T_{\it ref}$, which allows determining 
$\sigma_{\it ref}$ as 33~ps.
 The $\sigma_{pos}$ can be estimated to be 4.6~ps(=15.92~ps/12$^{1/2}$) 
by assuming $V_{\it eff}$ = 157~mm/ns (See section 5.2.2).

 Estimation of the timing resolution for a particular PMT at the counter 
center ($z = 0$) is then possible.
 That is, since the rms of the $T_{i}$ distribution, 
$\sigma_{T_{i}}$, was found to be 76~ps, the relation 
${\sigma_{i}}^{2}={\sigma_{T_{i}}}^{2}-{\sigma_{\it ref}}^{2}
-{\sigma_{pos}}^{2}$ allows 
determining the timing resolution per PMT, $\sigma_{i}$, as 68.3~ps.

\begin{figure}[hbtp]
  \begin{center}
    \includegraphics[width=6cm]{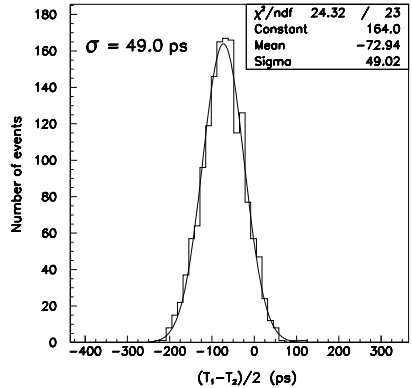}
  \caption
  [Timing distribution of $(T_{1}-T_{2})/2$ 
   at the counter center in a 1 GeV/$c$ $\pi^{+}$ beam.]
  {Timing distribution of $(T_{1}-T_{2})/2$ 
   at the counter center in a 1 GeV/$c$ $\pi^{+}$ beam.}
  \label{fig:aa_st_t12_pi}
  \end{center}
\end{figure}

 Estimation of the timing resolution per counter is similarly performed.
 Assuming that the timing resolution of $t_{1c}$ and $t_{2c}$ are the same 
at the counter center ($z = 0$), namely $\sigma_{1} = \sigma_{2}$, 
the beam crossing time for the counter at $z = 0$ is simply given as 
the mean time from Eq. (\ref{eq:pd3}), i.e.,
\begin{equation}
(T_{1}+T_{2})/2 = (t_{1c}+t_{2c})/2 - (\frac{L}{2})/V_{\it eff} - T_{\it ref} .
  \label{eq:crs}
\end{equation}
 From Eq. (\ref{eq:crs}) the timing resolution per counter, 
$\sigma(\frac{t_{1c}+t_{2c}}{2})$, can be estimated from the rms of the 
$(T_{1}-T_{2})/2$ distribution.
 As the $T_{\it ref}$ term subtracts out in 
$(T_{1}-T_{2})/2 (= (t_{1c}-t_{2c})/2 )$, 
the rms of $(T_{1}-T_{2})/2$ can be regarded as the timing 
resolution per counter at $z = 0$.
 From the rms of $(T_{1}-T_{2})/2$ for a pion beam, 
$\sigma(\frac{t_{1c}+t_{2c}}{2})$ is estimated to be 
49.0~ps (Fig.~\ref{fig:aa_st_t12_pi}).
 Since the rms of the distribution of $(T_{1}+T_{2})/2$, 
$\sigma(\frac{T_{1}+T_{2}}{2})$, includes $\sigma_{\it ref}$ and 
$\sigma(\frac{t_{1c}+t_{2c}}{2})$ in quadrature 
($(\sigma(\frac{T_{1}+T_{2}}{2}))^{2}=(\sigma(\frac{t_{1c}+t_{2c}}{2}))^{2}
+(\sigma_{\it ref})^{2}$), 
60.3~ps obtained from the rms $\sigma(\frac{T_{1}+T_{2}}{2})$ is consistent 
with $\sigma(\frac{t_{1c}+t_{2c}}{2}) = 49.0$~ps.

 Now, using the timing resolution per PMT, 68.5~ps at $z = 0$, the timing 
resolution per counter at $z = 0$ can also be estimated in another way.
 Since we assume that $\sigma_{1}$ and $\sigma_{2}$ are the same at $z = 0$, 
$\sigma(\frac{t_{1c}+t_{2c}}{2})$ can be expressed as
\begin{equation}
 (\sigma(\frac{t_{1c}+t_{2c}}{2}))^{2} 
           = (\frac{\sigma_{1}}{2})^{2}+(\frac{\sigma_{2}}{2})^{2} 
           = (\frac{\sigma_{1}}{\sqrt{2}})^{2} .
  \label{eq:int2}
\end{equation}
$\sigma(\frac{t_{1c}+t_{2c}}{2})$ is accordingly estimated to be 48.4~ps, 
which is consistent with 49.0~ps estimated from the rms of the 
$(T_{1}-T_{2})/2$ distribution.

 Apart from the $\pi^{+}$'s beam, we also estimate the timing resolution 
using a 1~GeV/$c$ proton beam.
 Since 1~GeV/$c$ proton has smaller $\beta(=0.73$) than 
that of 1~GeV/$c$ $\pi^{+}$, 
the d$E$/d$x$ is larger than that of $\pi^{+}$, 
resulting in better timing resolution.
 In a 1~GeV/$c$ proton beam, $\sigma(\frac{t_{1c}+t_{2c}}{2})$ from
$(T_{1}-T_{2})/2$ is similarly estimated to be 41.5~ps.

 Hereafter we discuss on the $z$-dependence.
 We studied the scale factor of $N_{pe}$ per ADC count in advance 
by using LED and electronics of beam test setup.
 Using the scale factor, $N_{\rm pe}$ per PMT 
is $\sim 500$ at the counter center for a minimum ionizing $\pi$.
 This value is substantially larger compared to 
BESS'95$(\sim 200)$~\cite{kn:det} due 
mainly to using a larger photocathode to improve light collection.
 The increase in $N_{\rm pe}$ is considered essential in the improvement
of timing resolution per counter from 78 to 50~ps.

\begin{figure}[hbtp]
   \hspace*{-1.0cm}
    \includegraphics[width=16cm]{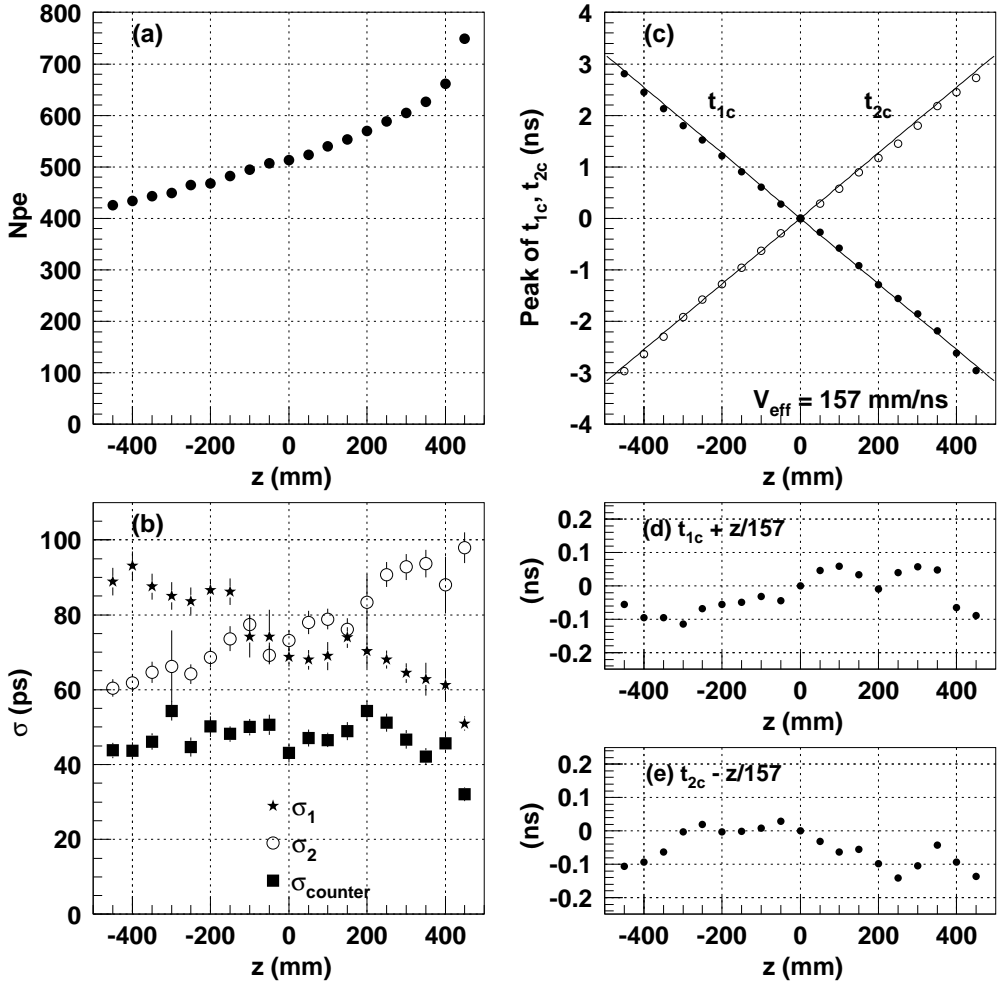}
  \caption
  [Dependence on $z$-position of (a) $N_{\rm pe}$, (b) $\sigma_{i}$, 
   (c) peak of $t_{1c}$, $t_{2c}$, and (d) and (f) timing difference 
   from $V_{\it eff}$ = 157~mm/ns.
   $V_{\it eff}$ represents typical effective velocity of 
   light in the scintillator.]
  {Dependence on $z$-position of (a) $N_{\rm pe}$, (b) $\sigma_{i}$, 
   (c) peak of $t_{1c}$, $t_{2c}$, and (d) and (f) timing difference 
   using $V_{\it eff}$ = 157~mm/ns, 
   where $V_{\it eff}$ is the typical effective velocity of 
   light in the scintillator.}
  \label{fig:tfpfig}
\end{figure}

 Figure~\ref{fig:tfpfig}(a) shows $N_{\rm pe}$ detected by PMT1 as a function 
of $z$, where $N_{\rm pe}$ is about 2 times larger at the nearest point than 
that at the furthest point.
 Corresponding plots for timing resolution of each PMT and TOF counter 
(by weighted average from Eq. (\ref{eq:pd3})) are shown in 
Fig.~\ref{fig:tfpfig}(b).
 Although a PMT's timing resolution improves as $N_{\rm pe}$ increases, 
ranging 50 to 100~ps, the resolution of the counter is 50~ps over
the entire scintillator length (95~cm) and is independent of the hit position.

\subsubsection{Effective propagation velocity of scintillation light}

 Figure~\ref{fig:tfpfig}(c) shows variations in $t_{1c}$ and $t_{2c}$ 
(Eq. (\ref{eq:slew1})) over $z$, and streight lines corresponding to 
$V_{\it eff}$ = 157~mm/ns.
 Figure~\ref{fig:tfpfig}(d) and (e) show the deviation(${\it diff(z)}$) of 
$t_{1c}$ and $t_{2c}$ from the timing calculated using $V_{\it eff}$ = 
157~mm/ns.
 The $z$-corrected timing determined with constant $V_{\it eff}$ 
does not show good linearity along $z$.
 When the timing resolution should be evaluated at any $z$-position, 
it is mandatory to correct for ${\it diff(z)}$.
 The ${\it diff(z)}$ for each counter shows the same tendency, 
but they are slightly different counter by counter.
Their differences seem to come from the wrapping condition of each counter.
 In a beam test, we confirmed 
the differences among several counters of the same type and 
the differences among counters with different wrapping materials.

\section{Performance Test with Cosmic-Rays}
 Using the full BESS'97 configuration (Fig.~\ref{fig:bess}), counter 
performance was evaluated with cosmic-ray data collected at ground-level at 
KEK during May 1997.
 That is, cosmic-ray muons were measured under a nominal central 
magnetic-field of 1 Tesla using the full electronics and trigger system of 
the BESS detector~\cite{kn:det}.
 We use 2~hours of data (215,014~events) to estimate TOF performance.
 An off-line selection criteria was applied to all events: (i) proper energy 
loss (d$E$/d$x$) is required to select particles with the electric charge 
$|Z|$ = 1 
in the upper and lower scintillator as well as in the JET 
chamber, and (ii) track quality cuts~\cite{kn:pbar93,kn:pbar95} are applied 
to ensure sufficient quality of tracks.

\begin{figure}[hbtp]
  \begin{center}
    \includegraphics[width=10cm]{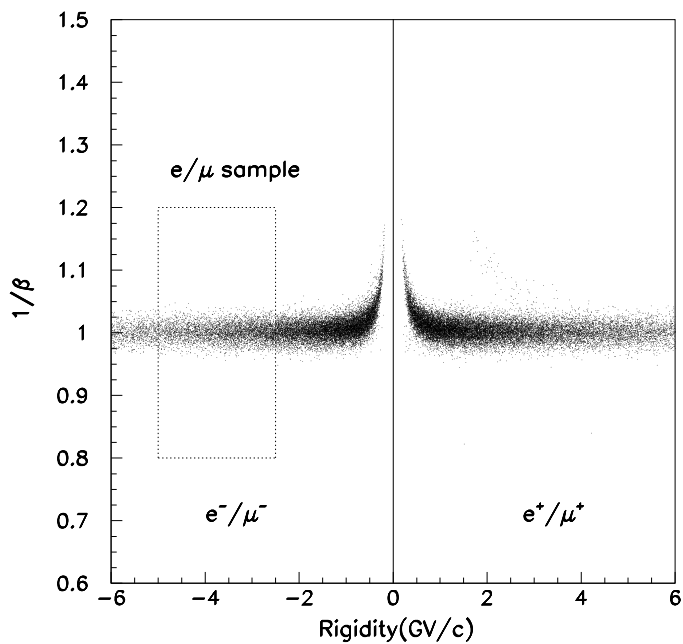}
    \caption
   [Scatter plot of 1/$\beta$ vs. rigidity using cosmic-ray data 
   taken at ground-level]
   {Scatter plot of 1/$\beta$ vs. rigidity for cosmic-ray data
    taken at ground-level using 
    cuts of d$E$/d$x$, track quality, and $1/\beta<1.2$.}
    \label{fig:gess973}
  \end{center}
\end{figure}

 The measured timing was corrected for time-walk correction, 
and $z$-dependent timing using ${\it diff(z)}$.
 To evaluate counter performance for particles with $\beta = 1$, 
negatively charged muons are selected in the rigidity 
range from --5 to --2.5 GV/$c$.
 Figure~\ref{fig:gess973} shows electron/muon ($e$/$\mu$) bands after
the cuts of d$E$/d$x$, track quality, and $1/\beta<1.2$, where the $e$/$\mu$ 
sample area is indicated (dashed rectangle).
 Since the selected sample mainly consists of muons, 
we treated all of them as muons.
 Based on the selected sample, timing resolution was determined by 
calculating $\Delta T$, which is the difference between the TOF obtained from 
the data of TOF PMTs, $T_{\it tof}$ (defined in section 4), and the TOF 
expected from the tracking information, $T_{trk}$, i.e.,
\begin{equation}
  \Delta T = T_{\it tof} - T_{trk} ,
  \label{eq:dtof1}
\end{equation}
\begin{equation}
  T_{trk} = \frac{L_{path}}{c\beta_{trk}(R,m)} = \frac{L_{path}}{c}\frac{E}{p} 
          = \frac{L_{path}}{c}\sqrt{ \frac{(ZR)^{2}+m^{2}}{(ZR)^{2}} } ,
  \label{eq:dtof3}
\end{equation}
where $L_{path}$ is the path length of the incident particles from upper 
to lower layer, and $c$ the velocity of light.
 Since the error in $R$ is relatively small, the error in $T_{trk}$ 
(Eq. (\ref{eq:dtof3})) is also negligible and the rms of $\Delta T$, being
71~ps which is consistent with the result of the beam test, 
therefore represents the resolution of the TOF hodoscope in the BESS detector.

\begin{figure}[hbtp]
  \begin{center}
    \includegraphics[width=7cm]{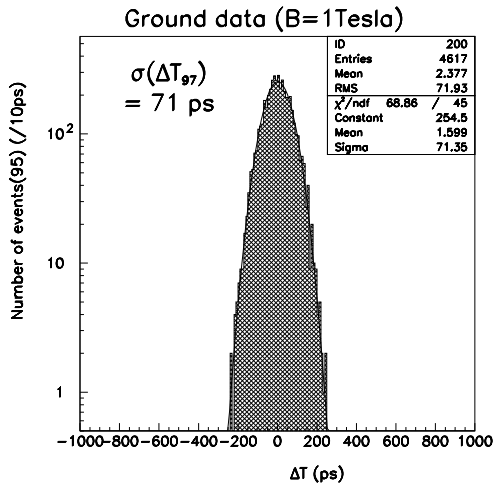}
    \caption
    [$\Delta T$ distribution for BESS '97 data at ground-level.]
    {$\Delta T$ distribution for BESS '97 data at ground-level.}
    \label{fig:gess97}
  \end{center}
\end{figure}

 Figure~\ref{fig:gess97} shows the $\Delta T$ distribution fitted to a pure 
Gaussian resolution function with no tail.
 Actually, however, since $T_{\it tof}$ is the time-of-flight between upper 
and lower TOF counters, each individual counter has a timing
resolution of 50~ps.

\section{BESS'97 vs. BESS'95 data}
\begin{figure}[hbtp]
 \begin{minipage}[t]{7cm}
  \begin{center}
    \includegraphics[width=7cm]{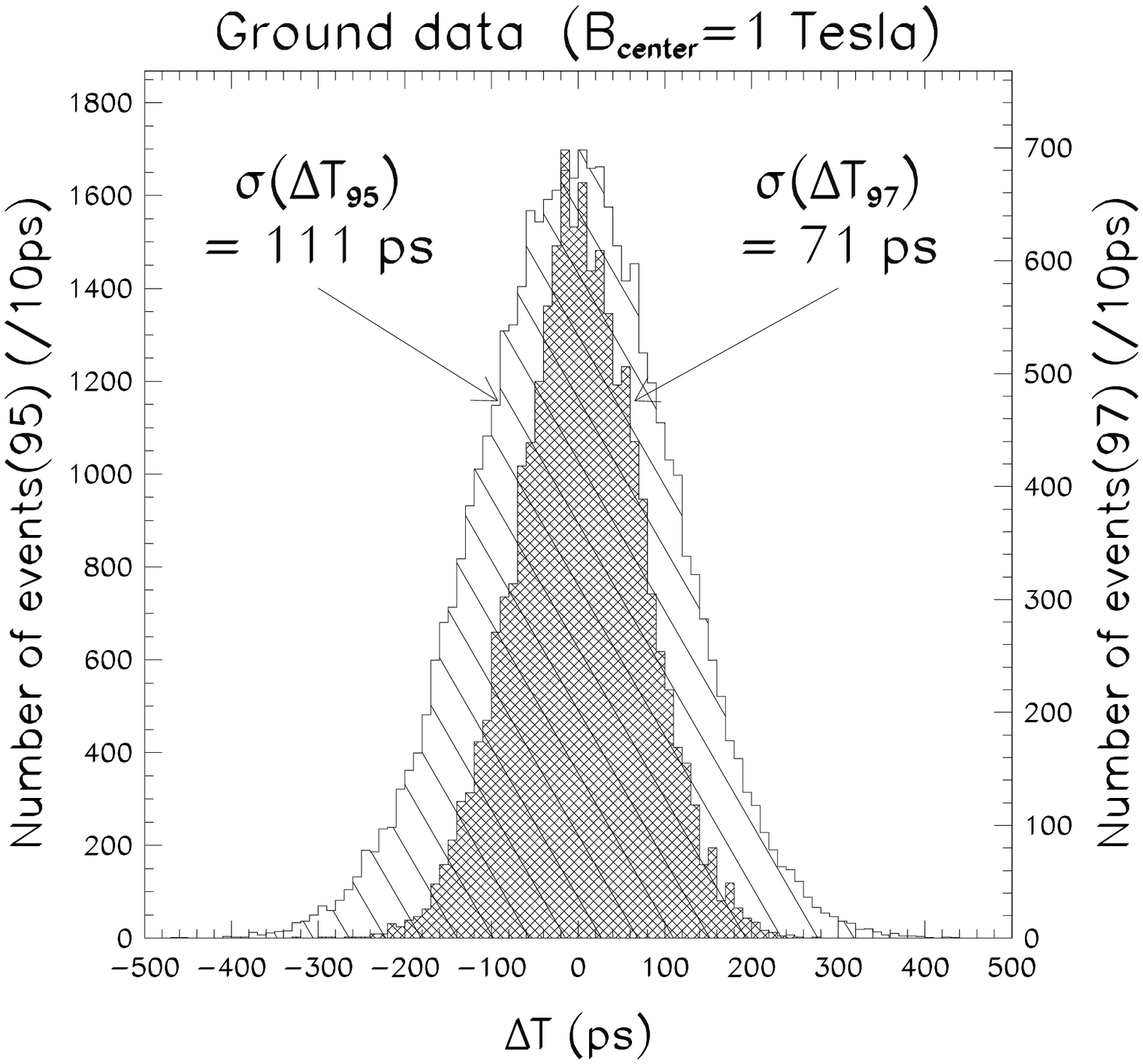}
    \caption
    [Comparison of $\Delta T$ resolution of BESS '97 and '95]
    {Comparison of $\Delta T$ resolution of BESS '97 and '95 
     using ground-level data.}
    \label{fig:dtof_g}
  \end{center}
 \end{minipage}\hspace{5mm}
 \begin{minipage}[t]{7cm}
  \begin{center}
    \includegraphics[width=7cm]{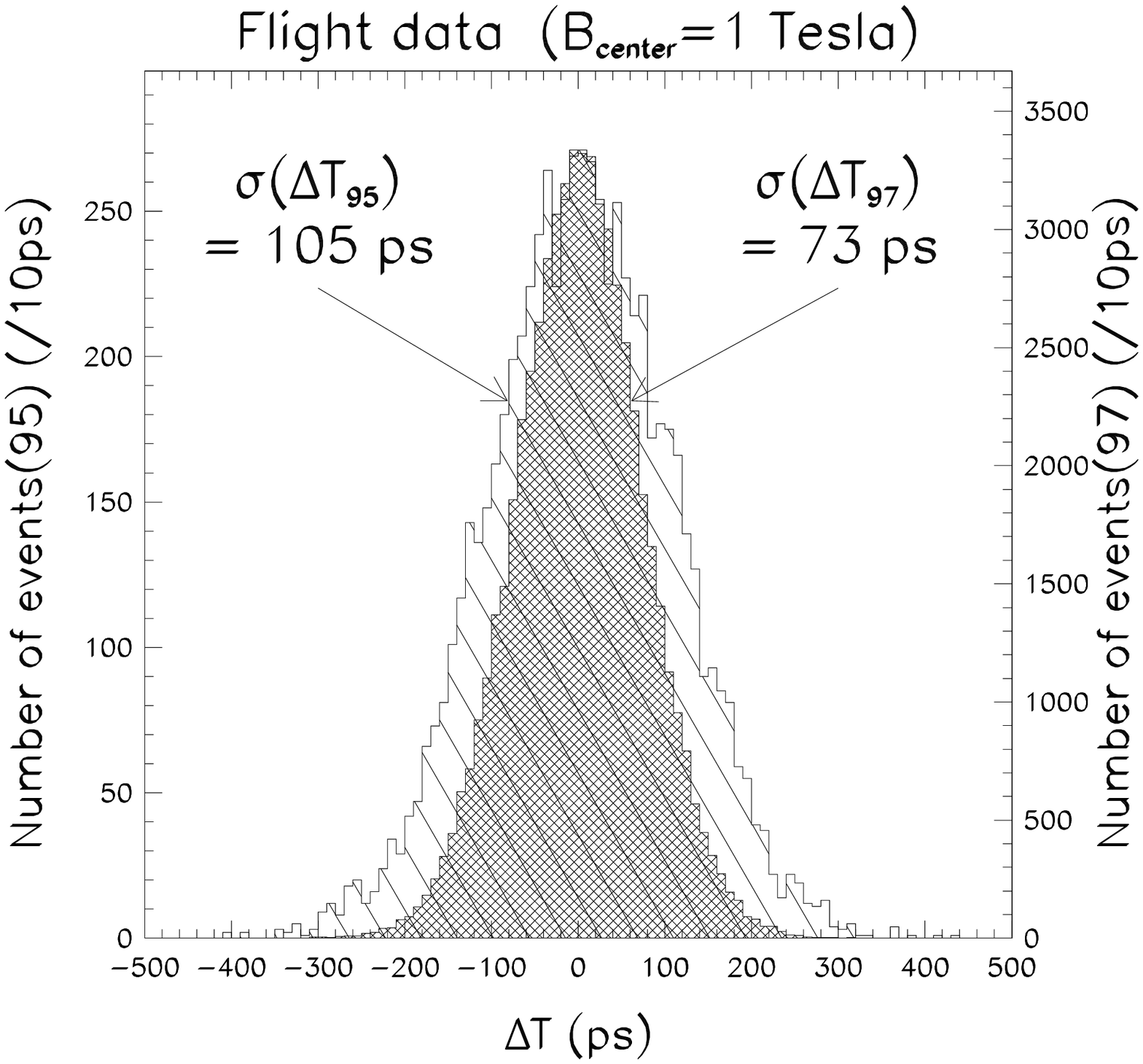}
    \caption
    [Comparison of $\Delta T$ resolution of BESS '97 and BESS '95]
    {Comparison of $\Delta T$ resolution of BESS '97 and '95 
     using data at an altitude of 36~km with 
     residual atmosphere of 5~g/cm$^{2}$.}
    \label{fig:dtof_a}
  \end{center}
 \end{minipage}
\end{figure}

\begin{figure}[hbtp]
  \begin{center}
    \includegraphics[width=7cm]{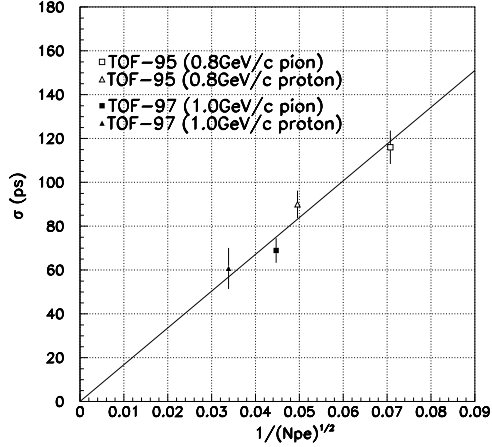}
    \caption
    [Linear relation between timing resolution and $1/\sqrt{N_{\rm pe}}$.]
    {Linear relation between timing resolution and $1/\sqrt{N_{\rm pe}}$.
     The solid line is a fit with a function of 
     $\sigma = a / \sqrt{N_{pe}}$, where $a=1679$~ps.}
    \label{fig:sn_comp}
  \end{center}
\end{figure}

 The improvement in TOF resolution from BESS'95 to BESS'97 is easily seen in 
Figs. \ref{fig:dtof_g} and \ref{fig:dtof_a} for ground 
(using the same muon sample as selected in section 6) and flight data 
(using proton sample with momentum more than 20~GeV/$c$), 
respectively, being consistent with the increase in $N_{\rm pe}$.
 The $N_{\rm pe}$ per PMT at counter center increased from 200 (BESS'95) to 
500 (BESS'97), with timing resolution being proportional to 
1/$\sqrt{N_{\rm pe}}$~\cite{kn:npe}.
 The linear relation between timing resolution per PMT 
and $1/\sqrt{N_{\rm pe}}$ is shown in Fig.~\ref{fig:sn_comp}, 
where the $N_{\rm pe}$ is the effective number of photoelectrons per one PMT.
 The data were taken both at the counter center in beam tests.
 The solid line in Fig.~\ref{fig:sn_comp} is a fit with a function of 
$\sigma = a / \sqrt{N_{\rm pe}}$, where $a=1679$~ps.
 The scintillators of the BESS'95 and BESS'97 TOF counters are the same size, 
and their light guides have similar shapes.
 As mentioned earlier , however, the use of a 2.5 inch vs. 2 inch PMT provides 
an effective photocathode area that is 2.0 times larger; a change that is 
mainly responsible for the higher $N_{\rm pe}$ value.

\section{Summary and conclusions}
 Described here are improvements in the BESS TOF hodoscope due to 
incorporating 2.5 inch PMTs along with optimization of light guide 
and the wrapping material.
 That is, the 13th dynode, developed for highly charged particles, exhibits 
good linearity and no saturation up to the range corresponding to $10^{4}$ 
times for minimum ionizing particle, the effective number of photoelectrons 
at the counter center increased from 200 to 500, and the timing resolution 
per TOF counter was enhanced from 78 to 50~ps as measured in beam tests.
 Performance of the complete detector configuration was also confirmed at 
ground-level under a magnetic field.
 This improvement in the TOF counter, together with the extension of the 
distance from the upper to lower layer of TOF counters, contributed to 
achieving better mass identification of charged particles.
 In particular, by utilizing the TOF hodoscope together with an aerogel 
\v{C}erenkov counter, the upper bound of the antiproton detectable kinetic 
energy range was raised from 1.4 to 3.5~GeV~\cite{kn:pbar97}, 
and the capability of separating light isotopes accordingly enhanced.

\begin{ack}
 Sincere thanks are given to T.~Sanuki for valuable help in manufacturing 
the high voltage power supply, to T.~Tanizaki for assistance in investigating 
saturation of the 13th dynode used for highly charged particles, 
to M.~Sasaki, M.~Fujikawa and T.~Murata for hodoscope installation, 
and to other members of BESS collaboration for such tremendous efforts.
 This work was supported by a Grant-in-Aid for Scientific Research from 
the Japanese Ministry of Education, Science and Culture.
 The analysis was performed using the computing facilities at ICEPP, 
University of Tokyo.
\end{ack}

\end{document}